\documentclass[pdflatex,sn-mathphys-num]{sn-jnl}%
\usepackage{graphicx}%
\usepackage{multirow}%
\usepackage{amsmath,amssymb,amsfonts}%
\usepackage{amsthm}%
\usepackage{mathrsfs}%
\usepackage[title]{appendix}%
\usepackage{xcolor}%
\usepackage{textcomp}%
\usepackage{manyfoot}%
\usepackage{booktabs}%
\usepackage{algorithm}%
\usepackage{algorithmicx}%
\usepackage{algpseudocode}%
\usepackage{listings}%
\usepackage{changepage}  

\begin{document}

\title[Article Title]{Factors That Influence the Adoption of AI-enabled Conversational Agents (AICAs) as an Augmenting Therapeutic Tool by Frontline Healthcare Workers: From Technology Acceptance Model 3 (TAM3) Lens - A Systematic Mapping Review}

\author{\fnm{Rawan} \sur{AlMakinah}}\email{ralmakinah@albany.edu}

\affil{\orgdiv{AI in Complex Systems Lab}, \orgname{University at Albany, SUNY}, \orgaddress{\street{1220 Washington Ave}, \city{Albany}, \postcode{12226}, \state{New York}, \country{USA}}}

\abstract{Artificial intelligent (AI) conversational agents hold a promising future in the field of mental health, especially in helping marginalized communities that lack access to mental health support services. It is tempting to have a 24/7 mental health companion that can be accessed anywhere using mobile phones to provide therapist-like advice. Yet, caution should be taken, and studies around their feasibility need to be surveyed. Before adopting such a rapidly changing technology, studies on its feasibility should be explored, summarized, and synthesized to gain a solid understanding of the status quo and to enable us to build a framework that can guide us throughout the development and deployment processes. Different perspectives must be considered when investigating the feasibility of AI conversational agents, including the mental healthcare professional perspective. The literature can provide insights into their perspectives in terms of opportunities, concerns, and implications. Mental health professionals, the subject-matter experts in this field, have their points of view that should be understood and considered. This systematic literature review will explore mental health practitioners' attitudes toward AI conversational agents and the factors that affect their adoption and recommendation of the technology to augment their services and treatments. The TAM3 Framework will be the lens through which this systematic literature review will be conducted.}


\maketitle

\section{Introduction}

The World Health Organization has reported that during the COVID-19 pandemic, there was a 25\% increase in anxiety and depression \cite{world_health_organization_covid-19_2022}. The unpredictable levels of the global mental health crisis due to the impact of the COVID-19 pandemic, especially on the psychological well-being of vulnerable groups, necessitates cross-disciplinary, inter-sectoral initiatives to develop timely preventive and therapeutic mental healthcare methods to prevent adverse physical, economic, and emotional outcomes \cite{alsagri_machine_2020, jacka_prevention_2014, lange_coronavirus_2021, sweeney_insights_2024, tsamakis_covid19_2021, vammen_emotional_2016, wu_prevalence_2021}. Internet accessibility, along with the rapid development of technologies, has paved the way for innovative approaches to mental healthcare \cite{daley_preliminary_2020}. Several studies have examined the effectiveness of employing artificial intelligence (AI) technologies as aiding tools to detect and proactively prevent mental health problems early and constantly help people in mental distress navigate their day. 

Abd-Alrazaq et al.'s scoping review on the potential of wearable AI devices for detecting anxiety and depression, for instance, offers great promise in terms of prescreening assessment \cite{abd-alrazaq_wearable_2023}. Moreover, Abilkaiyrkyzy et al. digital twin dialogue system for early mental illness detection, which utilizes pre-trained BERT models and fine-tuned on the E-DAIC dataset, accurately detects signs of mental health problems with 69\% accuracy and 84.75\% acceptability and usability \cite{abilkaiyrkyzy_dialogue_2024}. Another study by Shojaei et al. used a co-design approach with nine art therapists to explore the integration of AI in art therapy, finding that AI-assisted tools, such as a deck of cards paired with AI-generated images, could enhance therapeutic engagement \cite{shojaei_when_2024}. One of the findings of Sun et al.'s semi-structured interviews and co-design workshops with music therapists, which sought to elicit the opportunities and challenges of integrating musical AIs into the current music therapy treatment process, was that there is potential for applying musical AIs to individual therapy processes, and that was from the therapists' perspective \cite{sun_understanding_2024}.

All the studies above illustrate the potential of incorporating AI technologies into mental healthcare. Yet, stepping into a field whose implementation in mental health is still in its early stages  \cite{alimour_quality_2024} requires careful pursuit. The perspectives of different stakeholders, especially mental healthcare practitioners, who play a pivotal role in enhancing individuals' mental health, need to be addressed. This paper introduces a systematic mapping review to comprehend mental health practitioners' (therapists/psychiatrists) attitudes toward adopting AI-enabled conversational agents to augment their therapeutic processes. Through the lens of the Technology Acceptance Model 3 (TAM3) framework, we aim to categorize the elicited findings from the studies, find the gaps, and build an overarching guidance for the development of AI technologies that tackles the concerns of therapists.

\section{Technology Acceptance Model 3 (TAM3)}
The Technology Acceptance Model (TAM) is an information systems theory model and one of the most widely utilized theoretical models explaining an individual's intention to use new technology \cite{klaic_using_2020, silva_davis_2015}. Fred Davis developed the first iteration of the TAM in the 1980s \cite{lai_literature_2017, legris_why_2003}. The model proposes that when users are presented with a new technology, several factors influence their decision about how and when they will use it \cite{klaic_using_2020, silva_davis_2015}. The probability of system use, an indicator of system success, can be predicted by a user's intention to use the technology, which is shaped by two primary beliefs: (1) their perception of how useful the technology is (perceived usefulness (PU)) and (2) their perception of whether the technology is easy to use (perceived ease of use (PEOU)) \cite{klaic_using_2020, lai_literature_2017, silva_davis_2015}. 

Davis extended his original TAM to TAM2, incorporating additional factors such as subjective norms, output quality, results demonstrability, job relevance, and perceived ease of use as determinants of perceived usefulness \cite{klaic_using_2020, legris_why_2003}. Tested with longitudinal research designs, TAM2 aimed to provide a more comprehensive model of technology acceptance \cite{legris_why_2003}. Despite this enhancement, empirical studies using TAM and TAM2 have shown that these models consistently explain about 40\% of the variance in individuals' intention to use and subsequent use of technology \cite{klaic_using_2020, legris_why_2003}. However, the results remain somewhat inconsistent, indicating that significant factors influencing technology acceptance may still be missing from these models \cite{legris_why_2003}.

The determinants of perceived ease of use model, developed concurrently, introduced factors anchoring beliefs about technology, such as computer self-efficacy, anxiety, playfulness, and external control perceptions, with experience and voluntary use moderating perceived usefulness \cite{klaic_using_2020}. Recently, this model and TAM2 were integrated into the TAM3. This integrated model suggests that perceived usefulness is shaped by factors like output quality and relevance to user needs, while perceived ease of use depends on personal beliefs about skills, including self-efficacy and anxiety; both perceptions can be mediated by external factors such as experience and available support resources \cite{klaic_using_2020}.

\begin{figure}[h]
\centering
\includegraphics[width=1\textwidth]{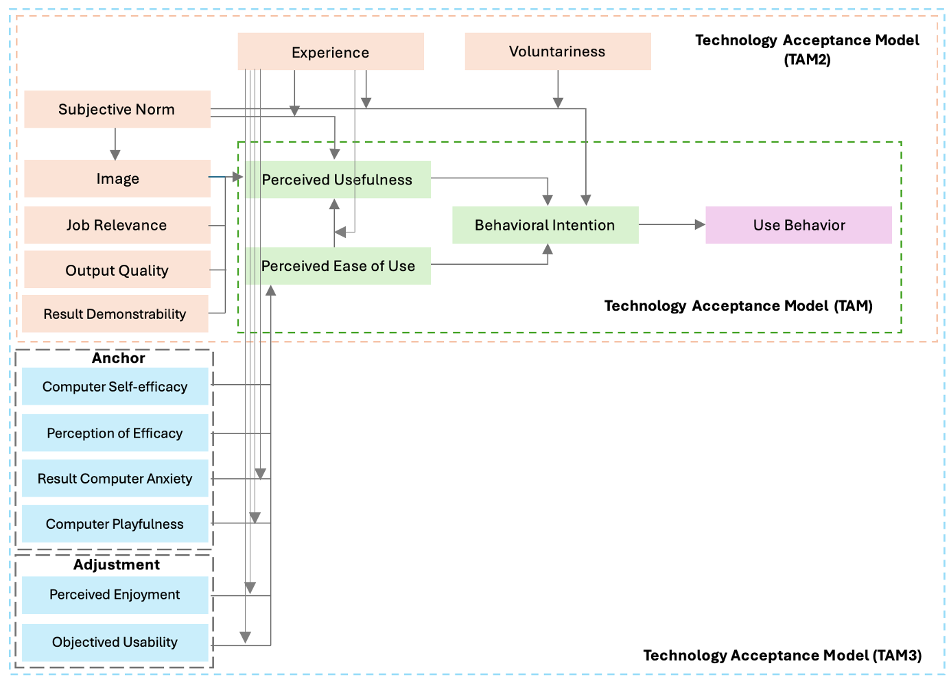}
\caption{Technology Acceptance Model 3 (TAM3) \cite{setiyani_using_2021}}\label{TAM3}
\end{figure}

TAM 3 includes 17 variables that act as influencing factors, as shown in Figure \ref{TAM3}. These variables are divided into anchor factors, including computer self-efficacy, perception of efficacy, computer anxiety, and computer playfulness, as well as adjustment factors, such as perceived enjoyment and objective usefulness. Other factors include image, job relevance, output quality, result demonstrability, subjective norm, experience, and voluntariness. These variables impact perceived usefulness and perceived ease of use, which in turn influence behavioral intention and use behavior \cite{setiyani_using_2021}. This model establishes a basis for further research aimed at understanding the factors influencing users' acceptance or rejection of information technology, as well as identifying ways to enhance its adoption.

The TAM3 model has been widely used in healthcare studies to evaluate factors influencing the acceptance and adoption of various health technologies by assessing perceived usefulness, perceived ease of use, and external moderating factors unique to healthcare environments. A study by Ho et al. examined factors affecting nurses' acceptance of a care plan system using TAM3 \cite{ho_determining_2020}. Data from 222 nurses in Taiwan showed that PU, PEOU, and subjective norms were key predictors of their intent to use the system, explaining 69\% of the variance. While PU and PEOU significantly impacted acceptance, moderating factors like voluntariness had no effect, likely due to the nurses' experience with the system.

Another study by Mwogosi and Kibusi evaluates Tanzanian healthcare practitioners' perceptions of EHR systems for clinical decision support in primary healthcare facilities \cite{mwogosi_effectiveness_2024}. Using TAM3 as a framework, the research found that while EHR systems are in use, their effectiveness is limited due to usability issues, lack of training, frequent crashes, and poor workflow integration. These challenges negatively impact user satisfaction, with frequent usage linked to lower satisfaction.

Moreover, Yazdanpanahi et al. conducted a study using TAM3 to identify factors influencing the adoption of teleorthodontic technology among Iranian orthodontists \cite{yazdanpanahi_investigating_2024}. Key findings showed that job relevance, output quality, and result demonstrability significantly impacted perceived usefulness, while perceived ease of use was mainly influenced by external control. Notably, subjective norms and personal image had minimal effect on behavioral intent. 

To the author's knowledge, and despite the growing literature using the TAM framework to predict user adoption of technology, none of the published studies focus on conducting a systematic literature review highlighting the adoption of AI-enabled conversational agents as augmenting tools for therapeutic processes in mental health from a therapist's perspective. This paper explores if published studies on AI in mental health can be mapped onto the variables in the TAM3 framework. The organization of this paper is structured as follows: The methodology section details the systematic mapping process, outlining the research questions, search and study selection procedures, and data extraction and analysis methods. The results section presents the findings from the mapping study and addresses the research questions. This is followed by the discussion section, which interprets the retrieved literature and provides context to the results. Finally, the conclusion section synthesizes the overall findings, highlighting the implications and potential future directions for research.

\section{Methodology}
This study integrates comprehensive literature into the TAM3 framework to explore and map mental health practitioners' attitudes toward adopting AI-enabled conversational agents into therapeutic processes as augmenting tools. We begin by extensively reviewing existing literature and academic articles to establish a foundational understanding. Following this, we identify key opportunities and concerns therapists developed while exploring and interacting with AI technologies. Next, we utilize these insights to construct the TAM3 framework that comprehensively categorizes and illustrates therapists' attitudes. 

\subsection{Literature Review}
A systematic mapping review was selected because the ultimate goal is to understand the literature on therapists' attitudes toward adopting AI-enabled conversational agents and categorize the findings using the TAM3 framework. This systematic mapping review will follow Petersen et al.'s guidelines \cite{petersen_guidelines_2015}. A systematic mapping review process is particularly advantageous when the topic area is broad and the quality and range of studies is diverse \cite{delgado_bias_2022}. The general process of systematic mapping review is to survey the research field, collect findings, classify them, and quantify the studies within established categories. We aim to address our specific research question within the mental health domain to uncover trends and identify knowledge gaps \cite{kabudi_ai-enabled_2021, petersen_guidelines_2015}.

A detailed description of systematic mapping studies follows. The first phase is the planning phase, which involves identifying the research question and deciding on the literature research strategy. The second phase is the execution phase, during which relevant papers are located through systematic searches, and the filtering and selection process is implemented based on titles and abstracts. Another name for filtering and selection is (keywording), meaning only relevant studies are considered. Subsequently, the selected articles undergo thorough screening for data extraction. The last phase synthesizes the collected data and maps it to a framework (TAM3 in our case) to provide a comprehensive overview of the research landscape.

\subsection{Research Question}
Following the guidelines for systematic mapping studies, this research aims to identify and categorize mental health practitioners' attitudes toward AI-enabled conversational agents as augmenting tools to therapeutic processes and construct a framework that illustrates and emphasizes the opportunities and concerning points to guide future development and implementation. This leads us to formulate two research questions:

\begin{itemize}
    \item \textbf{RQ1: What are healthcare professionals' attitudes and perceptions toward adopting AI technologies to augment their work?} This question will provide a comprehensive understanding of healthcare professionals' attitudes toward AI technologies overall. It might highlight opportunities, concerns, and implications of adopting AI technologies.
    \item \textbf{RQ2: What are frontline mental health professionals' attitudes and perceptions toward adopting AI-enabled conversational agents to augment therapeutic sessions?} This question will provide a specific understanding of therapists' attitudes toward AI-enabled conversational agents.
\end{itemize}

Narrowing the search scope may limit the exploration and understanding of therapists' attitudes. Additionally, a limited number of studies focus on mental health practitioners' attitudes toward AI-enabled conversational agents. Therefore, we will include studies that examine the attitudes of healthcare practitioners in general, as well as those explicitly addressing mental health practitioners.

\subsection{Search}
To answer the above research questions, a formal search was performed using an established database, namely \textit{Web of Science}. This database was selected for its extensive collections of peer-reviewed work, including recent academic journals and conference papers relevant to therapists' attitudes toward adopting AI technologies. Web of Science includes a significant number of AI-related publications, providing a comprehensive source of literature. The advanced search options in the Web of Science database allowed applying keywords and filters.

The search was limited to articles written in English and published within the last five years (2020–2024). This time frame was chosen to ensure that the review included the most current and relevant scholarly work, reflecting the rapid advancements and increasing attention to AI applications in the mental health field over recent years. A combination of general keywords and specific search criteria was employed to ensure inclusivity and avoid bias in the search results. Keywords used included terms like "Therapist", "Clinician", "counsellor", "counselor", "artificial intelligence", and "AI", with Boolean operators to incorporate synonyms and refine results. These keywords are listed in Table 1. 

\begin{table}[h]
\caption{Search Query}\label{}%
\begin{tabular}{@{}llll@{}}
\toprule
\textbf{Keywords} \\
\midrule
Therapist OR Clinician OR counsellor OR counselor OR   psychiatrist OR psychologist OR Psychotherapist \\ OR "social worker" OR "mental healthcare worker" OR "Mental health professional" OR \\ "Mental health Practitioners" OR "Cognitive behavioral therapist") AND ("artificial intelligence" OR AI \\ OR Chatbot OR "conversational engine" OR "conversational agent" OR "conversational artificial intelligence" \\ OR "AI-powered assistant" OR "Voice assistant" OR "AI-driven chatbot") AND (intervention OR treatment \\ OR recommend OR adopt OR Attitude OR Mediation OR Therapy OR care OR endorse OR implement OR \\ stance OR perspective OR viewpoint)  \\
\botrule
\end{tabular}
\end{table}

Figure 2 illustrates all articles retrieved from the Web of Science database. Two duplicate papers were manually identified and removed during the initial screening of the 54 papers retrieved. The titles and abstracts of the remaining 52 papers were then screened for eligibility, focusing on their evaluation of therapists' opinions toward AI technologies. Papers not meeting these criteria, as detailed in Table 2, were excluded, resulting in 20 excluded papers. We aimed to retrieve the full text of all relevant papers during the extraction process. We successfully obtained the full text for all 32 remaining papers. We thoroughly screened the entire paper to assess its eligibility and relevance to the topic under study, which led to the exclusion of five additional papers. Finally, we organized the remaining 27 papers into a table, extracting the summary of the studies and mapping the findings to the TAM3 framework.

\begin{figure}[h]
\centering
\includegraphics[width=1\textwidth]{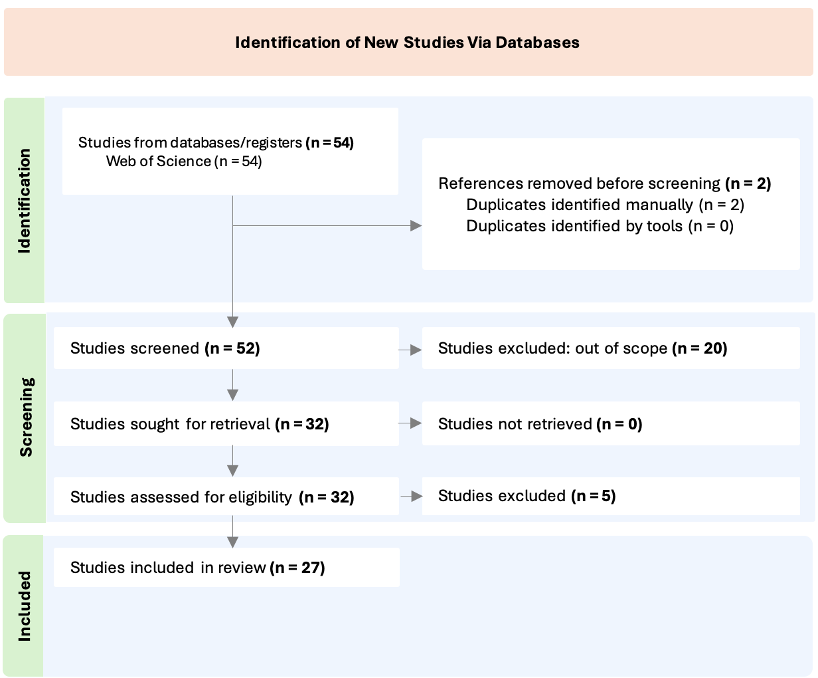}
\caption{Prisma flow diagram displaying the steps taken during the systematic literature review}\label{PRIZMA}
\end{figure}

\begin{table}[h]
\caption{Screening Publications for In/Exclusion Criteria}\label{tab1}%
\begin{tabular}{@{}llll@{}}
\toprule
\textbf{Criteria} & \textbf{Details}  \\
\midrule
\textbf{Inclusion Criteria}    & Interviews or surveys that elicit opinions or attitudes from mental health and medicine \\ 
& professionals.     \\
& Papers authored or co-authored by mental health and healthcare professionals.\\
\textbf{Exclusion Criteria}    & Studies evaluating AI vs. therapist assessment, judgment, or decision-making.\\
& Studies using AI as treatment decision support or in automatic patient care management.\\
& Studies suggesting AI use cases specifically for clinician benefit. \\
& Studies proposing design principles to enhance clinician adoption of AI. \\
& Experiments focusing on AI's role in clinician decision support, including safety comparisons. \\
& Studies evaluating AI as a clinician or embracing AI without considering the clinician's \\ 
& perspective. \\
& Papers discussing software developer responsibility in cases of clinician negligence with AI. \\
& Irrelevant fields where AI has another meaning or application. \\
& Unreliable sources (non-peer-reviewed papers, not articles). \\
& Studies on developing or improving AI models or AI-enhanced tools (e.g., robotic coaches). \\
& Studies using AI to determine treatment allocation (not perspective papers). \\
& Studies on user perspectives regarding AI as a counselor or treatment advisor. \\
& AI evaluation studies focused on diagnosis and treatment comparisons with clinicians or \\ & psychiatrists. \\
& Statistical studies comparing AI vs. clinicians in decision-making and treatment.\\
& Response Studies. \\
& Scoping / Mapping Reviews that do not include therapists' perspectives or attitudes. \\
\botrule
\end{tabular}
\end{table}

\section{Results}
Table 3 presents a summary of the systematic literature review conducted on the 27 papers identified. These papers explore the attitudes of mental health therapists toward AI-enabled conversational agents as augmentative tools for therapy sessions, as well as the perspectives of healthcare practitioners toward AI tools more broadly. The table outlines the general findings of each paper and maps these findings to the TAM3 framework. This process represents an initial step in the broader mapping effort, which aims to integrate all findings, align them with the TAM3 framework, identify common themes, and highlight existing gaps.

\section{Discussion}
This section synthesizes the findings from the literature on healthcare and mental health practitioners' attitudes toward AI technologies, including AI-enabled conversational agents (AICAs), through the lens of the Technology Acceptance Model 3 (TAM3). By analyzing the "Mapping to TAM3" column in Table 3, the discussion focuses on the acceptance of AI technologies, structured by key constructs from TAM3. The discussion addresses specific questions related to the determinants of perceived usefulness, ease of use, behavioral intention, use behavior, and moderating factors.

\subsection{AI Technologies and Perceived Usefulness}
\bmhead{Do healthcare and mental health practitioners perceive that AI technologies will be beneficial?} 
The degree to which practitioners believe that using AI technologies will enhance their job performance is critical to their acceptance. Studies reveal both opportunities and reservations regarding the perceived usefulness of AI technologies in healthcare and mental health practices. Practitioners generally perceive AI technologies as beneficial for improving efficiency, accessibility, and task management. For example, Carlbring et al. emphasized that conversational agents like ChatGPT can mimic empathy, reducing stigma and increasing accessibility in mental health care \cite{carlbring_new_2023}. Similarly, Doorn highlighted the usefulness of AI in automating alliance and symptom assessments in psychotherapy, although ethical concerns remain \cite{doorn_feasibility_2024}. However, tasks requiring deep emotional connection or complex decision-making reduce the perceived usefulness of these tools, as noted by Prescott and Hanley \cite{prescott_therapists_2023}.

\subsubsection{Determinants of Perceived Usefulness}
\textit{Subjective Norm (SN)}
\bmhead{Do practitioners feel social pressure from colleagues or the industry to adopt AI technologies?}
Subjective norms play a pivotal role in influencing the adoption of AI technologies among healthcare and mental health practitioners. Professional and organizational norms often emphasize ethical considerations and human oversight in diagnostics, as noted by Bienefeld et al. and Doorn \cite{bienefeld_human-ai_2024, doorn_feasibility_2024}. Clinicians' caution is shaped by industry standards prioritizing privacy and trust \cite{bradford_primary_2024}, while transparency concerns can affect patient acceptance when AI appears impersonal \cite{carlbring_new_2023}. Positive views on AI are frequently driven by peer influence and organizational endorsements, supported by regulatory bodies like the FDA and technology leaders, further enhancing clinician acceptance \cite{fazakarley_experiences_2023, ganapathy_artificial_2018}.\\ 

\textit{Image}
\bmhead{Do practitioners believe that adopting AI technologies will improve their professional image?}
Image was not discussed overtly in the reviewed studies, suggesting that healthcare and mental health practitioners may not prioritize professional image when adopting AI technologies. However, it can be inferred that AI could enhance perceived professionalism and innovation. For instance, Carlbring et al. implied that advanced AI tools boost the perceived quality of digital interventions, appealing to tech-savvy users \cite{carlbring_new_2023}. Fazakarley et al. further implied that AI might position healthcare providers as innovative, potentially motivating adoption \cite{fazakarley_experiences_2023}.\\

\textit{Job Relevance (JR)}
\bmhead{Do practitioners find AI technologies relevant to their professional tasks?}
Relevance depends on the specific applications of AI technologies. Carlbring et al. found that practitioners view these tools as relevant for structured tasks, such as symptom tracking and psychoeducational \cite{carlbring_new_2023}. However, Prescott and Hanley observed limited relevance in areas requiring strong emotional bonds and nuanced communication \cite{prescott_therapists_2023}.\\

\textit{Output Quality (OQ)}
\bmhead{Do practitioners perceive the outputs of AI technologies as high-quality and reliable?}
Practitioners' trust in the output quality of AI technologies often hinges on their reliability and accuracy in clinical contexts. Fazakarley et al. found that AI tools were valued for enhancing diagnostic accuracy and reducing errors, especially in data-intensive environments, such as continuous care settings \cite{fazakarley_experiences_2023}. Similarly, Ganapathy et al. demonstrated that AI-based systems achieved over 90\% accuracy in tasks like glioma grading, which increased trust among clinicians when the technology aligned with high-stakes medical decision-making \cite{ganapathy_artificial_2018}. However, Tikhomirov et al. highlighted concerns about AI systems misinterpreting irrelevant data, which could undermine their reliability and limit their applicability in nuanced, high-risk scenarios \cite{tikhomirov_medical_2024}.\\

\textit{Result Demonstrability (RD)}
\bmhead{Can practitioners clearly observe and demonstrate the benefits of using AI technologies?}
Result demonstrability is crucial for fostering trust and adoption of AI technologies among healthcare practitioners. Demonstrable benefits, such as improved diagnostic accuracy and efficiency, increase confidence in the tools \cite{ganapathy_artificial_2018}. In psychotherapy, tangible improvements in therapeutic alliance scores and symptom tracking validate the utility of AI systems, even as ethical concerns persist \cite{carlbring_new_2023, doorn_feasibility_2024}. However, skepticism remains when AI outputs are not easily interpretable, underscoring the need for transparent and user-friendly systems \cite{jones_artificial_2023, lam_potential_2024}. 

\subsection{AI Technologies and Perceived Ease of Use}
\bmhead{Do mental health therapists perceive that AI-enabled conversational agents are easy to use?}
The degree to which practitioners perceive AI technologies as easy to use is influenced by individual confidence, organizational support, and system design.

\subsubsection{Determinants of Perceived Ease of Use}
\textit{Computer Self-Efficacy (CSE)}
\bmhead{Do practitioners feel capable of effectively using AI technologies in their practice?}
The confidence of healthcare practitioners in their ability to use AI technologies plays an important role in their adoption. Training and education are essential \cite{fazakarley_experiences_2023, shawli_physical_2024} in building this confidence, as claimed by Keane and Topol, who highlighted the need for hands-on learning with public datasets to familiarize practitioners with AI's potential and limitations \cite{keane_ai-facilitated_2021}. However, Sebri et al. noted that age would require more supportive training to overcome hesitancy \cite{sebri_artificial_2020}. Therefore, structured and accessible training opportunities are crucial for fostering self-efficacy and facilitating broader adoption.\\

\textit{Perceptions of External Control (PEC)}
\bmhead{Do practitioners believe they have the necessary support and resources to use AI technologies?}
Perceptions of external control play a pivotal role in the adoption of AI technologies among healthcare practitioners. Studies emphasize the importance of organizational support, such as clear implementation guidelines, in reducing barriers to adoption \cite{shawli_physical_2024}. Additionally, practitioners express concerns about data regulations, limited resources, resource strain, and hospital systems \cite{fazakarley_experiences_2023}, highlighting the need for robust systems to ensure seamless integration into clinical workflows. \\

\textit{Computer Anxiety (CA)}
\bmhead{Do practitioners experience anxiety when considering the use of AI technologies?}
Practitioners often experience anxiety about adopting AI technologies due to fears of job displacement, concerns over data security, or uncertainty about system reliability. Sebri et al. highlighted apprehension among therapists about AI's ability to handle nuanced therapeutic tasks, contributing to feelings of unease \cite{sebri_artificial_2020}. Similarly,  Doorn and Shawli et al. noted that ethical and privacy concerns further exacerbate hesitation, particularly for those unfamiliar with advanced technology \cite{doorn_feasibility_2024, shawli_physical_2024}. To mitigate these concerns, training and organizational support have been recommended as key strategies to boost confidence and reduce anxiety. \\

\textit{Computer Playfulness (CP)}
\bmhead{Do practitioners find experimenting with AI technologies to be enjoyable and engaging?}
The concept of Computer Playfulness (CP), which reflects practitioners' ability to engage with and explore AI technologies in an enjoyable and spontaneous way, remains underexplored in the provided studies. However, Cunningham et al. highlighted that interactive AI tools capable of providing real-time feedback could enhance engagement by encouraging therapists to explore empathy-mimicking capabilities \cite{cunningham_opening_2023}. Yet, more evidence is needed to fully understand the role of playfulness in AI adoption across healthcare settings. \\

\textit{Perceived Enjoyment (PE)}
\bmhead{Do practitioners enjoy using AI technologies in their practice?}
Perceived enjoyment of AI technologies among healthcare and mental health practitioners depends on their ability to reduce repetitive tasks and introduce engaging functionalities. Doorn highlighted that interactive AI systems can enhance job satisfaction by streamlining routine activities and enabling creative engagement in therapy workflows \cite{doorn_feasibility_2024}. However, Sebri et al. noted limited enjoyment or enthusiasm for AI across groups under study, reflecting low desirability, except for cognitive behavioral therapists, who appreciated AI's alignment with structured therapeutic interventions \cite{sebri_artificial_2020}. This suggests that while interactive and well-designed AI tools can increase job satisfaction for some practitioners, enthusiasm remains tempered by perceived limitations in AI's ability to address complex tasks. \\

\textit{Objective Usability (OU)}
\bmhead{Are AI technologies designed in a way that makes them inherently easy to use for practitioners?}
Objective usability refers to the actual effort required to use AI technologies, determined by system design and functionality. Carlbring et al. highlighted that tools like ChatGPT can simplify therapeutic workflows, but their usability depends on their ability to integrate seamlessly into clinical processes \cite{carlbring_new_2023}. Keane and Topol emphasized the importance of intuitive, automated AI tools that do not require extensive technical knowledge, making them accessible even to non-coders \cite{keane_ai-facilitated_2021}. Sivaraman et al. pointed out that visual explanations and user-friendly interfaces can enhance usability, although impractical features like fixed dosage scales in critical care settings can reduce efficiency \cite{sivaraman_ignore_2023}. Usability challenges, such as those noted by Tikhomirov et al., arise when AI systems fail to account for the contextual reasoning that clinicians rely on, leading to limited practical adoption \cite{tikhomirov_medical_2024}.

\subsection{AI Technologies and Behavioral Intentions}
\bmhead{Are healthcare and mental health practitioners motivated or willing to exert the effort to incorporate AI technologies into their practice?}
Behavioral intention toward adopting AI technologies varies across healthcare domains. Many practitioners recognize the potential of AI to enhance efficiency and decision-making, particularly in structured and repetitive tasks, such as diagnostics \cite{lam_potential_2024, shawli_physical_2024}. However, skepticism persists regarding AI's reliability and its ability to address nuanced, human-centered interactions, as noted by Sebri et al., where practitioners doubted AI's ability to replicate emotional understanding and interpret the true meaning of unconscious states or systemic relationships \cite{sebri_artificial_2020}. Additionally, concerns about accountability and legal implications influence hesitation in adoption, as highlighted by Jones et al. \cite{jones_artificial_2023}. These barriers are often compounded by ethical concerns about data privacy and transparency, which clinicians view as critical for trust \cite{fazakarley_experiences_2023, tikhomirov_medical_2024}. 

\subsection{Technologies and Use Behavior}
\bmhead{How are healthcare and mental health practitioners using AI technologies?}
Practitioners primarily use AI technologies in supplemental or assistive roles rather than core therapeutic processes. For example, AI tools are often deployed for monitoring, triage, and administrative tasks like documentation, as highlighted by Sivaraman et al., where ICU clinicians used AI selectively for treatment recommendations but avoided over-reliance due to concerns about accuracy and alignment with clinical judgment \cite{sivaraman_ignore_2023}. In primary care, Holt-Quick noted the integration of AI in administrative tasks, such as streamlining workflows, but this often required humans in the loop (HITL) due to accountability concerns \cite{holt-quick_primary_2024}. These findings indicate that current use behaviors are cautious and predominantly focused on areas where AI complements human expertise rather than replacing it. 

\subsection{Moderating Factors}
\subsubsection{Experience}
\bmhead{How does prior experience with technology influence therapists' perceptions and use of AI-enabled conversational agents?}
Prior experience with technology significantly shapes practitioners' attitudes toward AI adoption. Kleine et al. noted that healthcare practitioners with greater familiarity and training in AI are more confident in integrating such tools into their practice, leading to higher adoption rates \cite{kleine_attitudes_2023}. In the context of physical therapy, Shawli et al. highlighted that exposure to AI in educational or professional settings improved acceptance, particularly among practitioners with postgraduate qualifications or over a decade of experience \cite{shawli_physical_2024}. These findings indicate that technical familiarity and positive prior exposure can foster trust in AI technologies.

\subsubsection{Voluntariness}
\bmhead{Does the optional or mandatory nature of adopting AI technologies affect practitioners' acceptance and use?}
The reviewed literature provides limited insight into the role of voluntariness in the adoption of AI technologies by healthcare and mental health practitioners. None of the 27 papers explicitly address whether the optional or mandatory nature of AI adoption impacts practitioners' acceptance and use. While some studies, such as Shawli et al., emphasize the importance of guidelines and flexibility, the broader implications of voluntariness as a moderating factor remain underexplored \cite{shawli_physical_2024}. This represents a gap in the existing body of research, highlighting the need for future studies to investigate how the perceived autonomy of adoption influences the acceptance and integration of AI technologies in healthcare settings.

\section{Implications and Future Research}
The findings from the discussion provide several critical implications for practice and highlight areas requiring further exploration in the adoption of AI technologies among healthcare and mental health practitioners.

\subsection{Implications for Practice}
The integration of AI technologies in healthcare and mental health practices presents a unique opportunity to enhance efficiency, accessibility, and decision-making. However, the findings indicate that practitioners remain cautious, particularly in areas requiring emotional sensitivity and complex reasoning. This calls for the following practical actions:

\begin{itemize} 
    \item \textbf{Training and Education:} Structured training programs tailored to varying levels of computer self-efficacy are essential. Training should focus on enhancing practitioners' understanding of AI capabilities, addressing ethical concerns, and building confidence in using AI technologies effectively \cite{fazakarley_experiences_2023, keane_ai-facilitated_2021, shawli_physical_2024}.

    \item \textbf{Organizational Support:} Institutions must address perceptions of external control by providing technical infrastructure, clear implementation guidelines, and adequate IT support. This includes ensuring that systems are intuitive, secure, and aligned with clinical workflows \cite{fazakarley_experiences_2023, shawli_physical_2024}.
    
    \item \textbf{Addressing Ethical Concerns:} Ethical considerations, including data privacy and the risk of over-reliance, should be central to AI adoption strategies. Transparent governance frameworks and clinician oversight in critical decision-making processes are crucial to building trust and reducing anxiety \cite{bienefeld_human-ai_2024, carlbring_new_2023, doorn_feasibility_2024, shawli_physical_2024, tikhomirov_medical_2024}. Additionally, the skepticism surrounding AI's role in emotionally nuanced tasks highlights the need for balanced human-AI collaboration. AI tools should complement rather than replace human expertise, particularly in high-stakes scenarios such as psychotherapy and critical care \cite{prescott_therapists_2023}.
\end{itemize}

\subsection{Directions for Future Research}
Using the Technology Acceptance Model 3 (TAM3) as a framework, several critical gaps in the literature have been identified. While many factors influencing the adoption of AI technologies among healthcare and mental health practitioners have been explored, some key determinants remain underexamined, requiring focused attention in future research:

\begin{itemize} 
    \item \textbf{Computer Playfulness (CP):} The concept of playfulness, which involves spontaneous and enjoyable interactions with AI technologies, remains largely unaddressed in the reviewed studies. While Cunningham et al. suggested that interactive tools could enhance engagement, no empirical evidence specifically examines how playfulness influences adoption behavior or long-term use \cite{cunningham_opening_2023}. Future research should investigate how playful interactions with AI systems might mitigate anxiety, improve self-efficacy, and foster greater experimentation and acceptance.

    \item \textbf{Voluntariness:} The reviewed literature fails to adequately explore the role of voluntariness in AI adoption. Specifically, it is unclear how optional versus mandatory use impacts behavioral intentions and perceived ease of use. This gap is significant, as perceptions of autonomy could directly influence practitioner satisfaction and motivation to adopt AI technologies. Future research should investigate the interplay between voluntariness, perceived usefulness, and organizational policies in driving acceptance. 
    
    \item \textbf{Image:} The concept of Image remains insufficiently addressed in the reviewed literature regarding AI adoption among healthcare and mental health practitioners. No study explicitly investigates the degree to which practitioners perceive AI technologies as tools that enhance their professional status or reputation. Future research should explore how the perception of AI's ability to improve image influences adoption, particularly in environments where maintaining professional credibility and innovation are crucial. Additionally, studies could examine whether this perception varies across healthcare roles, levels of experience, or regions, providing a deeper understanding of how Image impacts behavior.
\end{itemize}

\begin{figure}[h]
\centering
\includegraphics[width=1\textwidth]{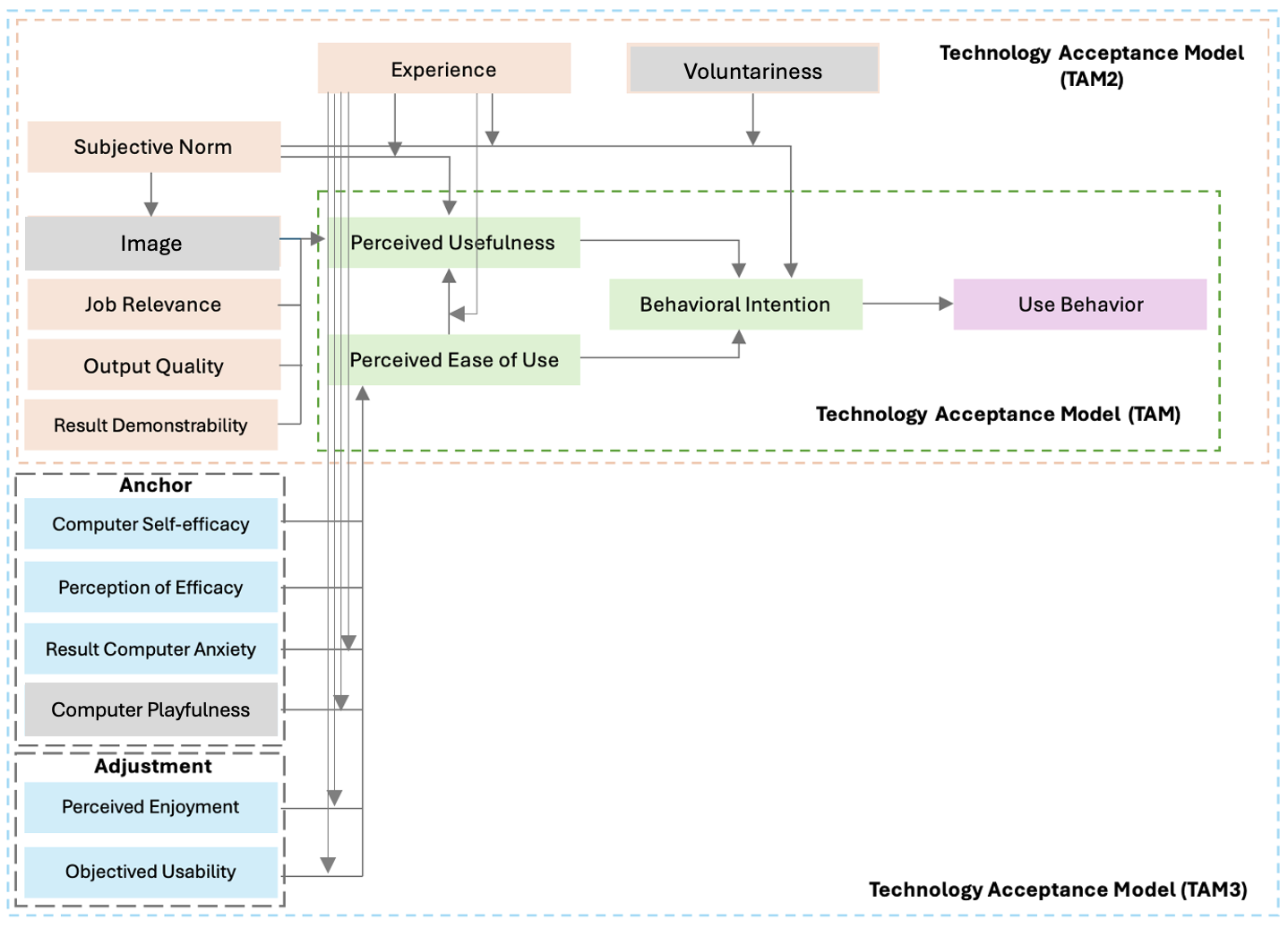}
\caption{TAM3 with the literature Gap (in Gray)}\label{Gap}
\end{figure}

\section{Limitation}
This study provides valuable insights into healthcare and mental health practitioners' attitudes toward AI technologies using the Technology Acceptance Model 3 (TAM3). However, several limitations should be noted:

\begin{enumerate}[1.]
    \item\textbf{Author Subjectivity in Filtering Studies:} Filtering studies for inclusion in the systematic mapping review involved a degree of author subjectivity. Decisions on which studies to include may have been influenced by the authors' interpretations, potentially affecting the comprehensiveness and objectivity of the review.
    \item\textbf{Subjectivity in Eliciting and Mapping Attitudes:} Mapping practitioners' attitudes to TAM3 factors required subjective judgment in interpreting and categorizing findings from the literature. While efforts were made to align findings systematically, the mapping may reflect biases in interpreting study results.
    \item\textbf{Database Limitation:} The systematic mapping review was conducted using only one database, Web of Science. Although this database is reputable and comprehensive, using additional databases, such as PubMed, Scopus, or IEEE Xplore, could enhance the breadth and depth of the review. Future studies should expand the database scope to ensure a more exhaustive literature review.
    \item\textbf{Exploration of Google Scholar:} Google Scholar was explored to identify potential similar studies, but no comparable research was found. While this supports the claim of novelty, the inherent variability in Google Scholar's indexing and search functionality may have limited the discovery of relevant studies.
    \item\textbf{Claim of Novelty:} The claim that this study is the first systematic application of TAM3 to analyze healthcare and mental health practitioners' attitudes toward AI technologies was framed as "to the authors' knowledge." This acknowledges the possibility that similar studies may exist but were not identified within the scope of this review.
\end{enumerate}

\section{Conclusion}
The integration of AI-enabled conversational agents into mental health practices holds significant potential to address the growing demand for accessible and efficient mental health support. However, this review highlights the importance of understanding the perspectives of healthcare practitioners, particularly mental health professionals, when designing and implementing such technologies. By using the Technology Acceptance Model 3 (TAM3) as a framework, this study systematically categorized practitioners' attitudes, identifying both opportunities and concerns regarding AI adoption.

Key findings indicate that while AI technologies are perceived as useful for routine and structured tasks, skepticism persists regarding their application in emotionally nuanced or high-risk clinical scenarios. Concerns about ethical implications and system reliability underscore the need for transparent design and robust support systems. Moreover, gaps in understanding factors such as voluntariness, playfulness, and professional image highlight areas requiring further research to foster adoption.

Ultimately, a balanced approach that prioritizes human-AI collaboration, practitioner training, and organizational support will be critical to achieving meaningful integration of AI technologies in mental health care. Future studies should continue to explore these dimensions to ensure the development of AI systems that align with both practitioner needs and ethical standards, paving the way for more effective and sustainable mental health interventions.

\section*{Declarations}
\bmhead{Funding} Not applicable
\bmhead{Conflict of interest / Competing interests} Not applicable
\bmhead{Ethics approval and consent to participate} Not applicable
\bmhead{Consent for publication} Not applicable
\bmhead{Data availability} Not applicable
\bmhead{Materials availability} Not applicable
\bmhead{Code availability} Not applicable

\bibliography{sn-TAM3}

\newpage
\begin{appendices}
\pagenumbering{gobble} 
\section{Retrieved Studies' Findings Summary and mapping to TAM3}\label{Mapping}

\begin{figure}[h]
\centering
\includegraphics[width=1\textwidth]{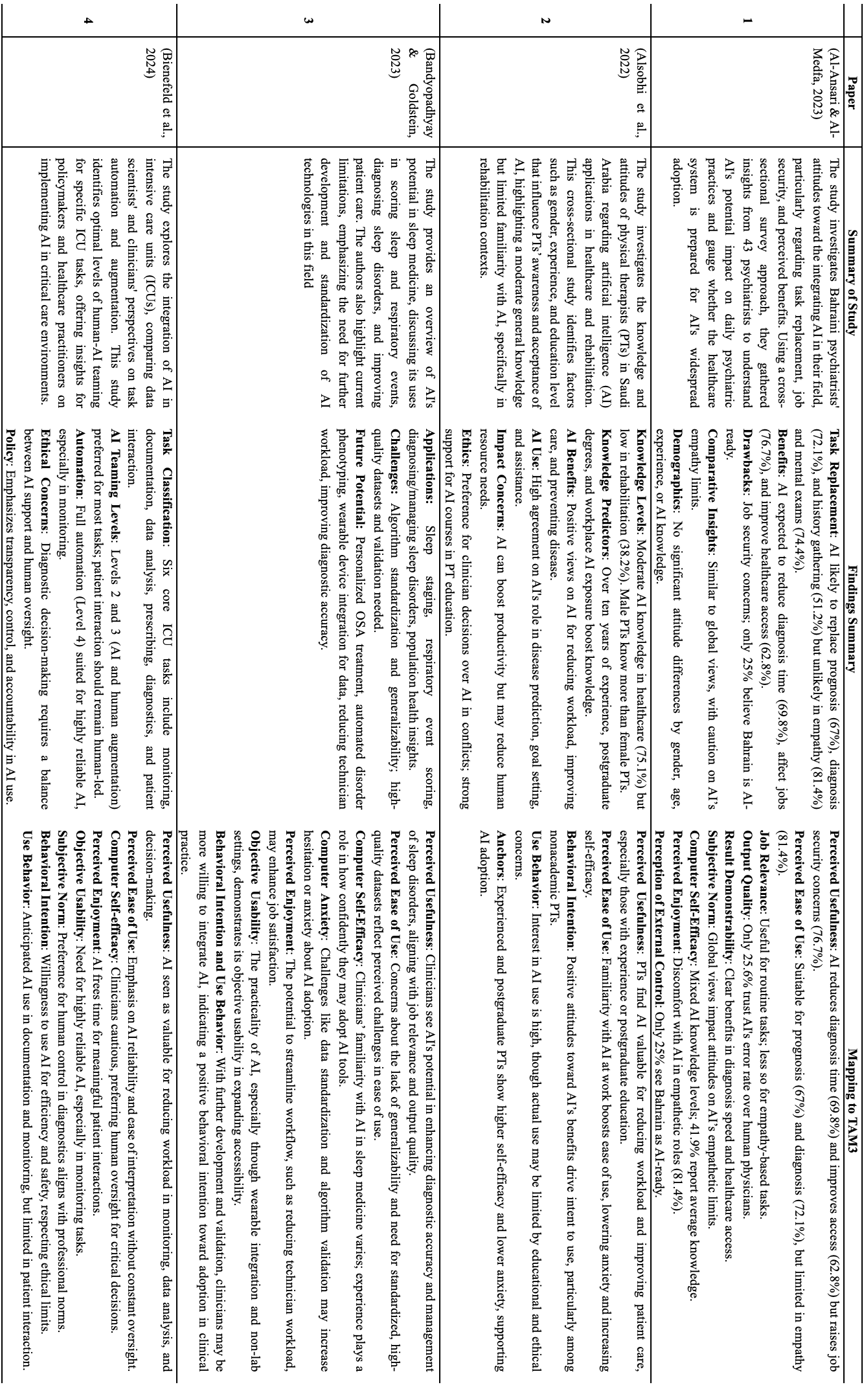}
\end{figure}

\begin{figure}[h]
\centering
\includegraphics[width=1\textwidth]{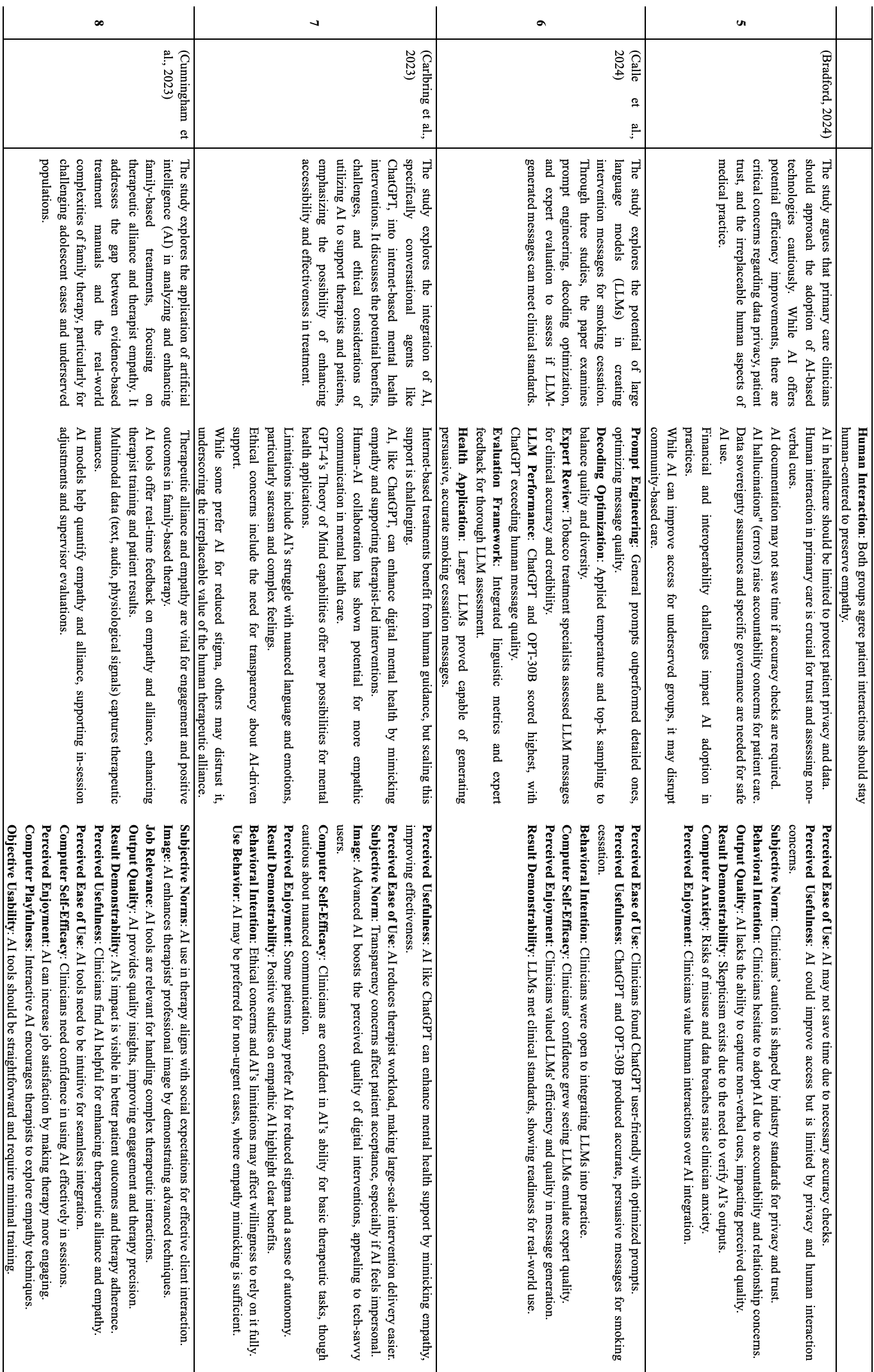}
\end{figure}
\begin{figure}[h]
\centering
\includegraphics[width=1\textwidth]{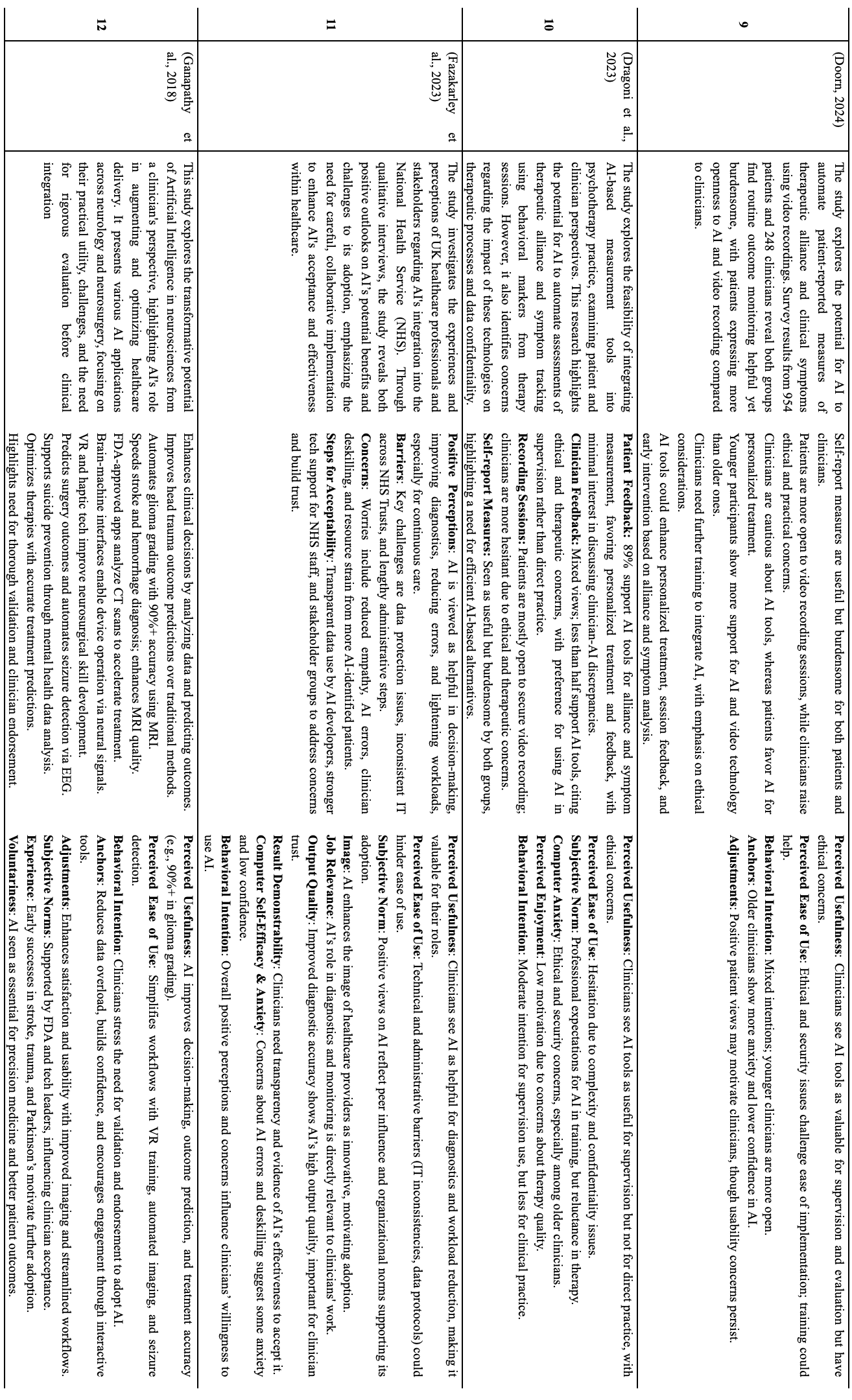}
\end{figure}

\begin{figure}[h]
\centering
\includegraphics[width=1\textwidth]{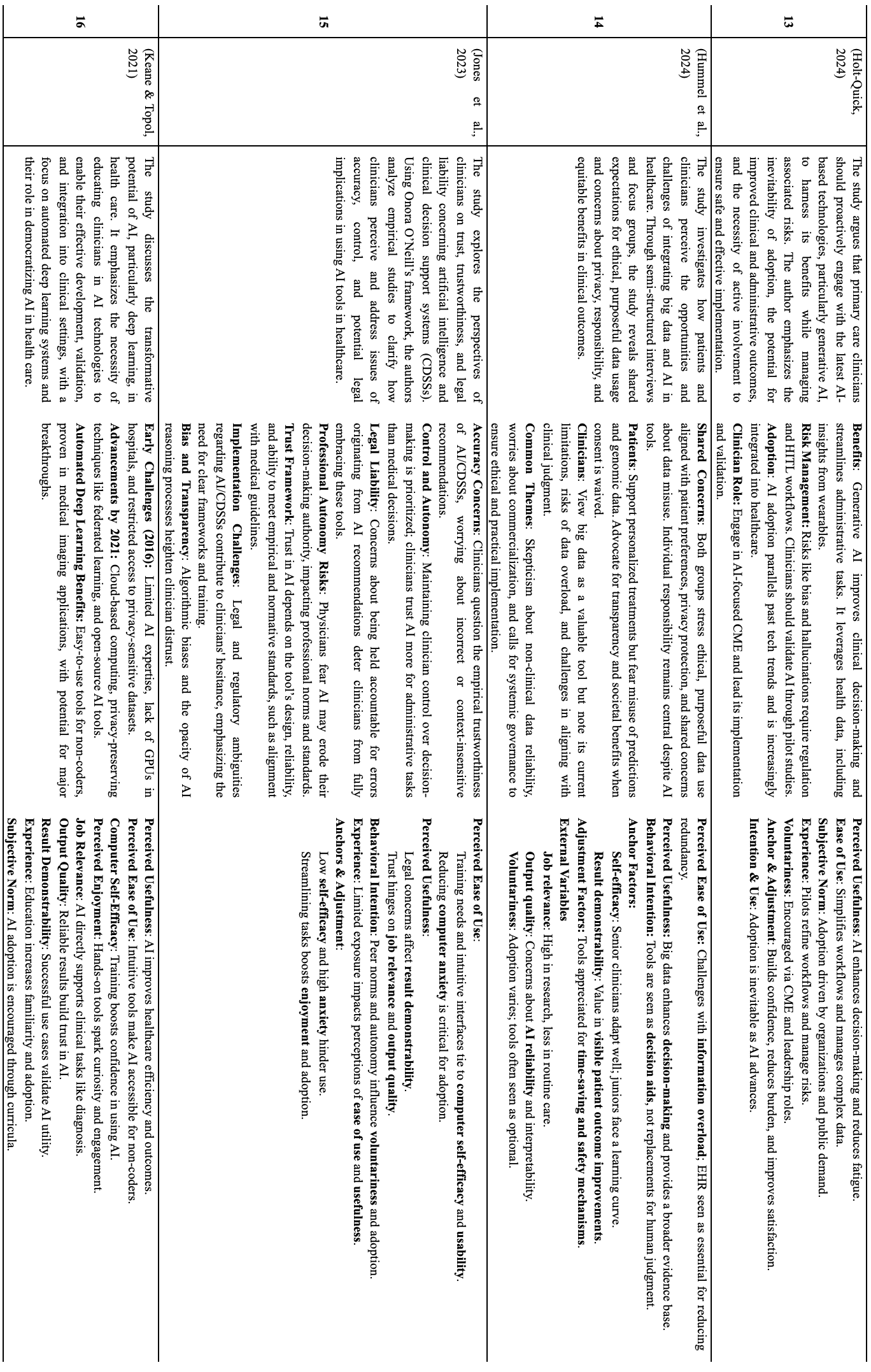}
\end{figure}

\begin{figure}[h]
\centering
\includegraphics[width=1\textwidth]{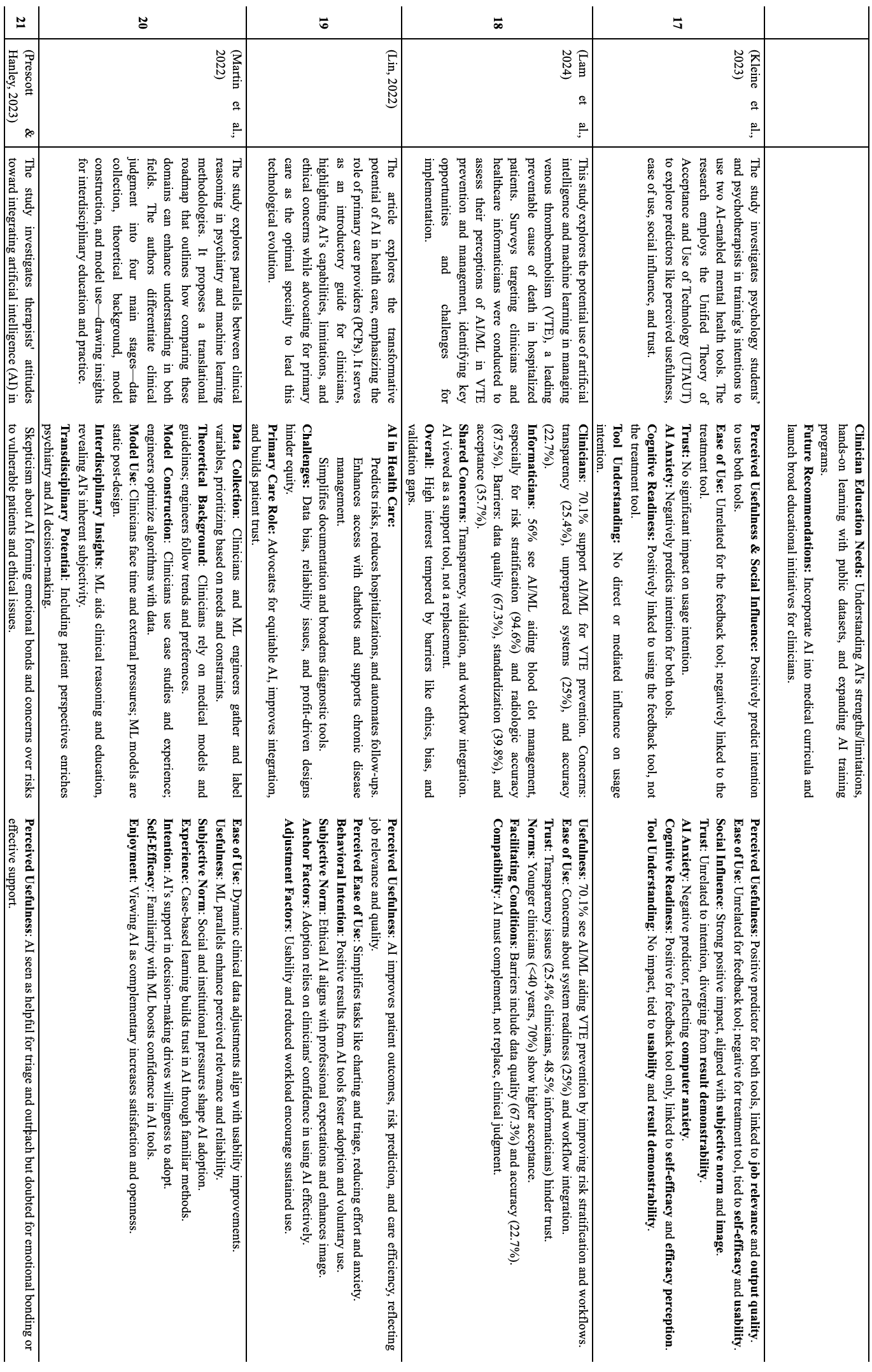}
\end{figure}

\begin{figure}[h]
\centering
\includegraphics[width=1\textwidth]{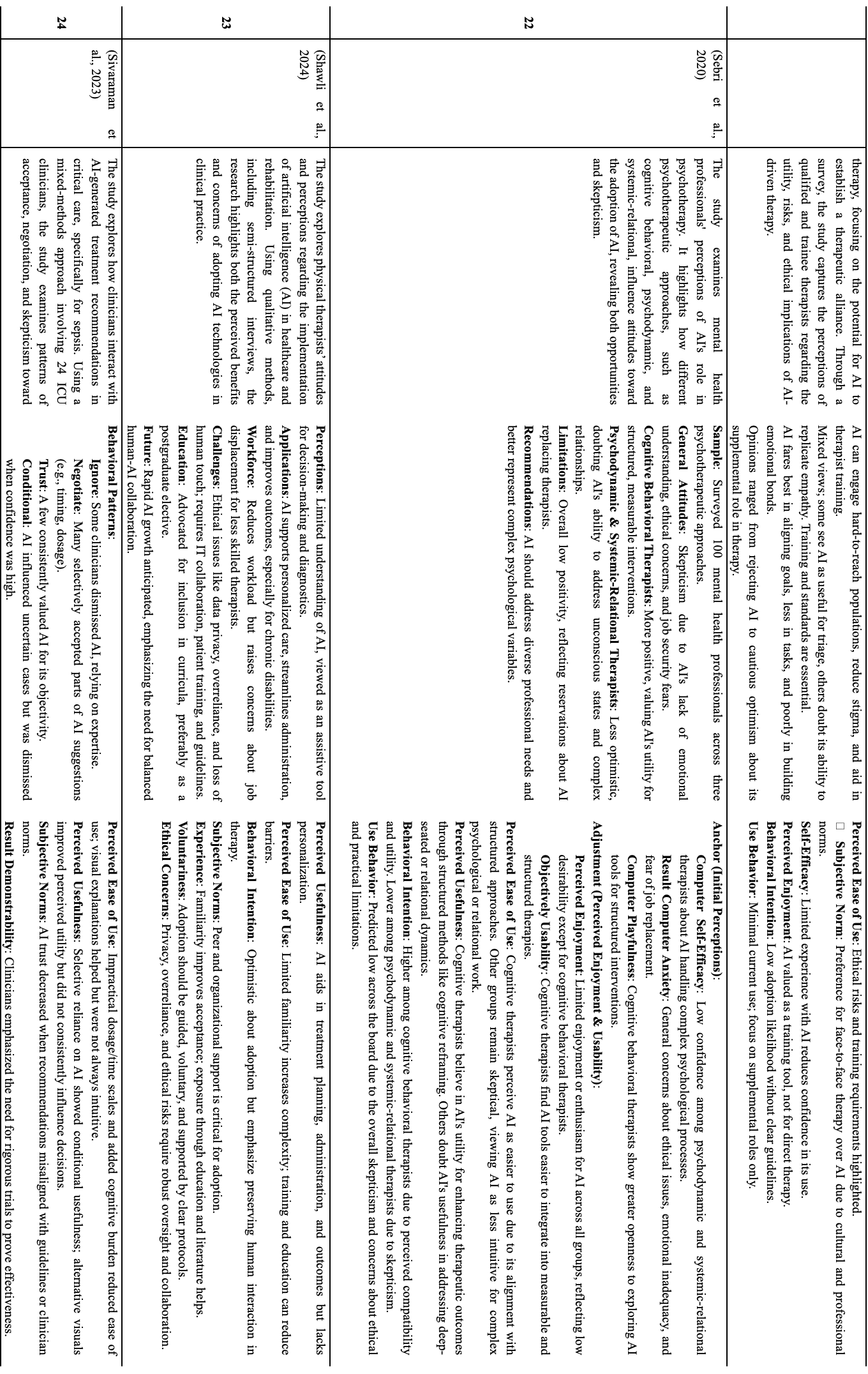}
\end{figure}

\begin{figure}[h]
\centering
\includegraphics[width=1\textwidth]{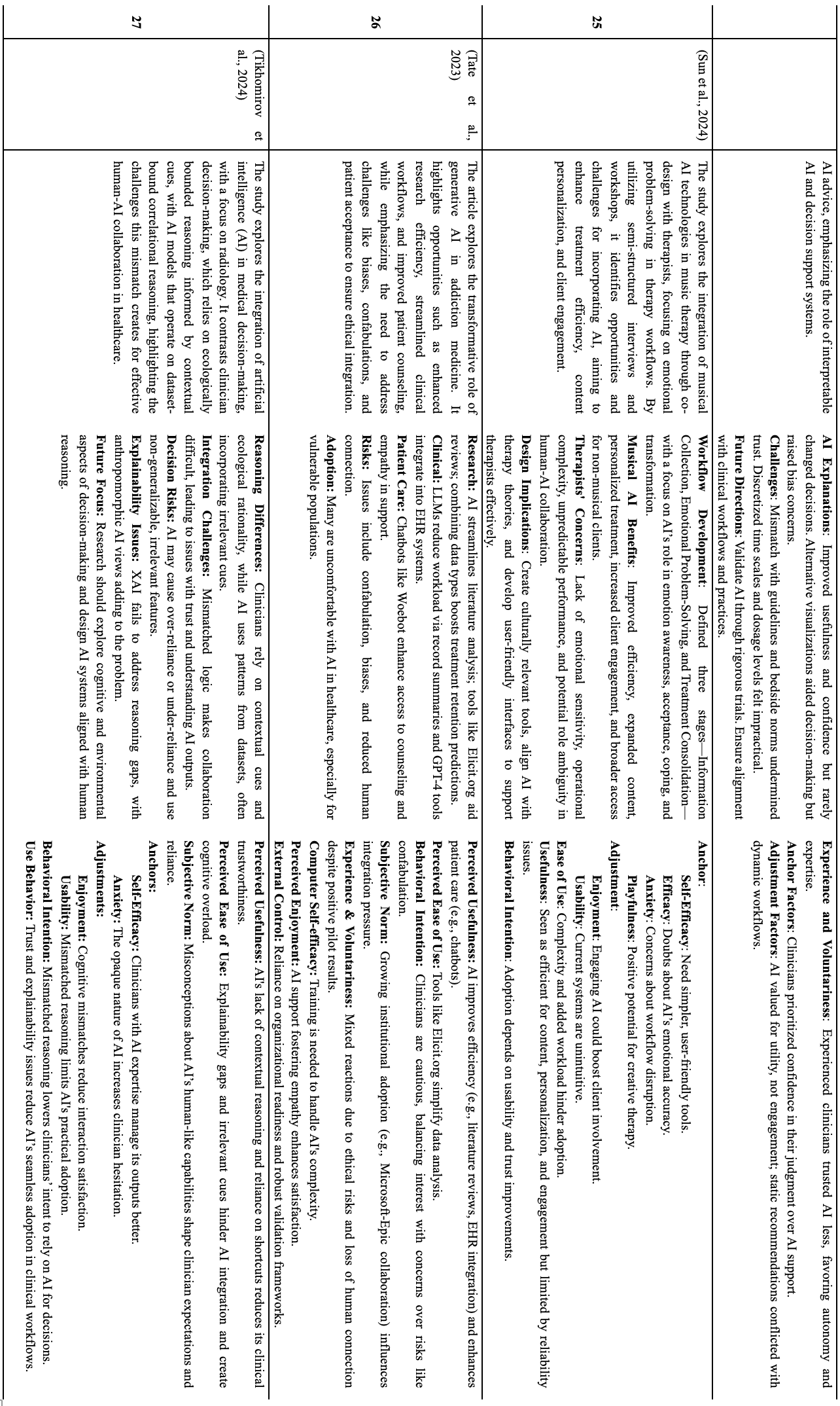}
\end{figure}

\end{appendices}

\end{document}